\documentclass[journal=jacsat,manuscript=article]{achemso}

\usepackage[version=3]{mhchem} 
\usepackage{xcolor}
\usepackage{hyperref}
\usepackage{amssymb}

\author{Junhan Chang}
\affiliation[Unknown University]
{College of Chemistry and Molecular Engineering, Peking University}
\alsoaffiliation[DP Tech]
{DP Technology}
\author{Duo Zhang}
\affiliation[DP Tech]
{DP Technology}
\alsoaffiliation[AISI]
{AI for Science Institute, Beijing}
\author{Yuqing Deng}
\affiliation[DP Tech]
{DP Technology}
\author{Hongrui Lin}
\affiliation[DP Tech]
{DP Technology}
\author{Zhirong Liu}
\affiliation[Unknown University]
{College of Chemistry and Molecular Engineering, Peking University}

\author{Linfeng Zhang}
\affiliation[DP Tech]
{DP Technology}
\alsoaffiliation[AISI]
{AI for Science Institute, Beijing}

\author{Hang Zheng}
\email{zhengh@dp.tech}
\affiliation[DP Tech]
{DP Technology}
\author{Xinyan Wang}
\email{wangxy@dp.tech}
\affiliation[DP Tech]
{DP Technology}

\title[An \textsf{achemso} demo]
  {Efficient and Precise Force Field Optimization for Biomolecules Using DPA-2}

\abbreviations{IR,NMR,UV}
\keywords{American Chemical Society, \LaTeX}

\begin{document}







\begin{abstract}
Molecular simulations are essential tools in computational chemistry, enabling the prediction and understanding of molecular interactions and thermodynamic properties of biomolecules. 
However, traditional force fields face significant challenges in accurately representing novel molecules and complex chemical environments due to the labor-intensive process of manually setting optimization parameters and the high computational cost of quantum mechanical calculations. 
To overcome these difficulties, we fine-tuned a high-accuracy DPA-2 pre-trained model and applied it to optimize force field parameters on-the-fly, significantly reducing computational costs. Our method combines this fine-tuned DPA-2 model with a node-embedding-based similarity metric, allowing seamless augmentation to new chemical species without manual intervention.
We applied this process to the TYK2 inhibitor and PTP1B systems and demonstrated its effectiveness through the improvement of free energy perturbation calculation results. 
This advancement contributes valuable insights and tools for the computational chemistry community.
\end{abstract}

\section{Introduction}


Molecular simulations play a crucial role in computational chemistry, offering powerful tools for predicting and understanding molecular interactions and the thermodynamic properties of biomolecules. 
Achieving accurate results in these applications requires extensive sampling of the conformation space over long timescales, ranging from nanoseconds to milliseconds. This extensive sampling is necessary to capture the dynamic behavior of molecules and ensure that the simulations reflect realistic biological conditions.

Traditional methods like \emph{ab initio} calculations, which are based on quantum mechanics, provide high accuracy but are computationally prohibitive for large systems and long timescales due to their high computational costs. Neural network potentials (NNPs)\cite{deepmd-prl,schütt2017schnet,ani2x,qiao2020orbnet,nequip,kovács2023maceoff23} offer similar accuracy with better efficiency but still require substantial computational resources for extensive simulations\cite{2020nnpmm,chodera2024nnpfep}.

To balance computational efficiency and accuracy, researchers often use force fields. Molecular force fields allow for faster simulations by approximating the interactions between atoms using predefined parameters.
Among the various molecular force fields developed to support these calculations, a few stand out for their widespread use and ongoing development to include a wide range of drug-like molecules: GAFF\cite{gaff,gaff-abcg2}, CGenFF\cite{cgenff,cgenff2016halogen}, OpenFF\cite{openff1.0,openff2.0} and OPLS\cite{opls-aa,opls3,opls3e,opls4}. Each force field has been carefully updated and expanded over time to capture better the diversity of drug-like molecules and their interactions with biological targets, reflecting the ongoing commitment within the scientific community to enhance the tools available for drug discovery and development. 




Force field optimization can be broadly classified into two categories: optimization of intermolecular terms and intramolecular terms. Optimization of intermolecular terms deals with the interactions between molecules, including electrostatic, repulsion, and dispersion interactions, and there is ongoing progress in this area\cite{bcc1,bcc2,resp,cgenff2016halogen,gaff-abcg2,opls3,opls4}. In contrast, optimization of intramolecular terms focuses on forces within a molecule, such as bond stretching, angle bending, and torsion profiles, and lacks a universal solution\cite{2020nnpmm}.

Optimization of intramolecular terms typically involves acquiring dihedral parameters through torsion scans, requiring extensive \emph{ab initio} calculations, and constructing a parameter library by manually defining numerous atom types or substructure types. This process is labor-intensive and difficult to generalize to novel molecular species.

To address these limitations, this study introduces a novel on-the-fly force field optimization approach. 
Firstly, we fine-tuned the DPA-2 pre-trained model \cite{zhang2023dpa2}, for precise potential surface prediction, then applied it to the torsion scan process, achieving significant computational acceleration. 
Secondly, we developed a node-embedding-based similarity metric to replace hand-crafted atom or substructure types. This allows for seamless augmentation to any new chemical species without manual intervention, thereby ensuring parameter consistency and reducing systematic errors.

By combining these techniques, our on-the-fly force field optimization approach addresses the key challenges of high computational expense and labor-intensive parameter library construction, paving the way for more efficient and broadly applicable force field optimization.


\section{Methods}
\label{sec:ff-strategy}

The primary objective of optimizing the intramolecular interaction terms in the force field is to ensure that the calculated energies for bond stretching, angle bending, and torsion profiles closely align with the results obtained from high-accuracy \emph{ab initio} calculations. Achieving this alignment enhances the precision of molecular simulations, thereby improving the accuracy of predictions related to molecular behavior. 
Previous studies \cite{openff1.0} and our experiments in \nameref{sec:inconsistency} have shown that quantum mechanical (QM) and molecular mechanics (MM) potential energy surfaces often exhibit different local minima. Fitting in rotamer space can mitigate the impact of these inconsistencies. 

The manual design of force field parameters is labor-intensive and challenging to extend to new scenarios. Recent advancements leverage methods such as graph neural networks to identify atoms with similar chemical environments, providing a more efficient and scalable solution\cite{wang2022espaloma}.

\begin{figure}[H]
    \centering
    \includegraphics[width=1.0\linewidth]{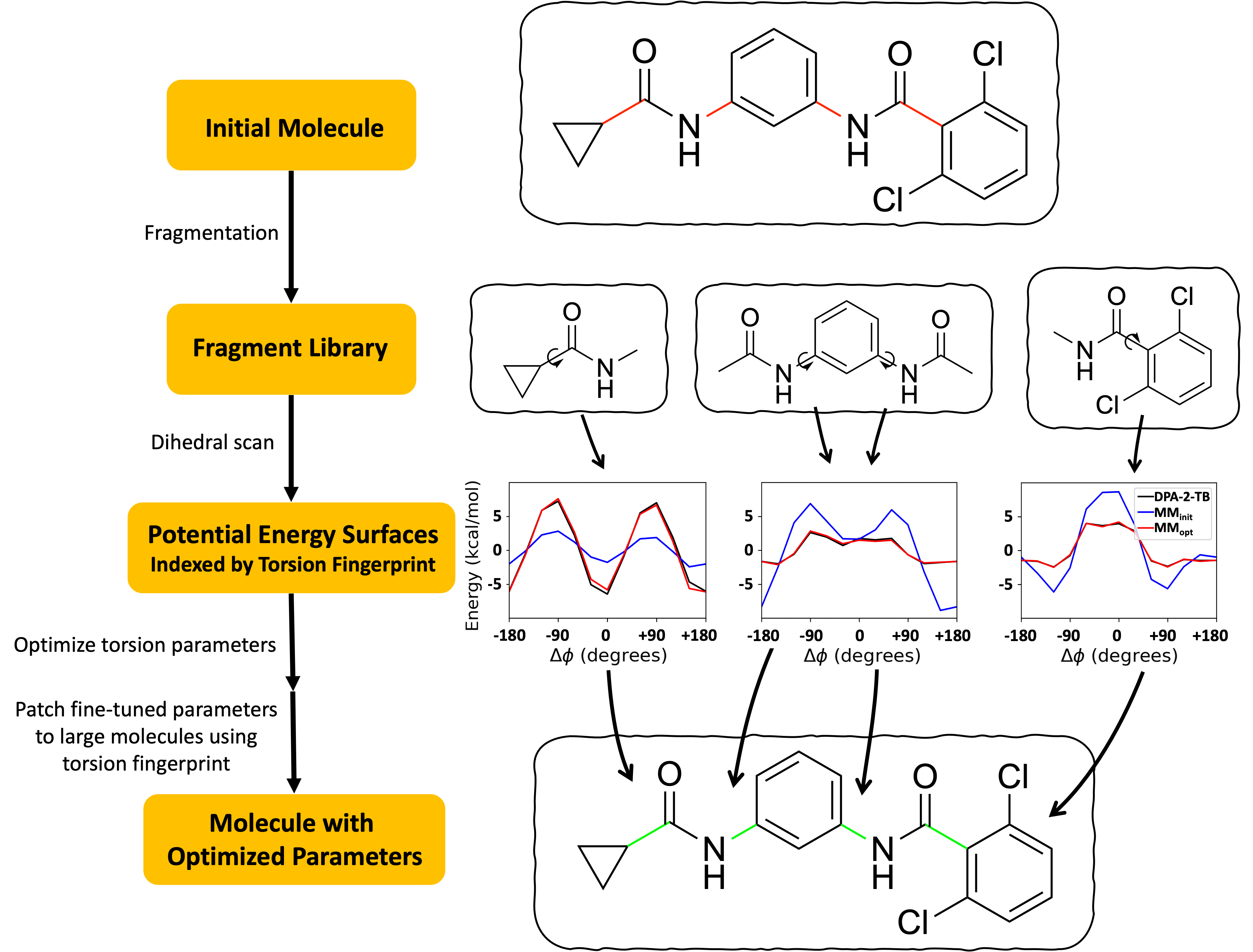}
    \caption{The workflow of the dihedral parameter optimization tool for molecular force fields.}
    \label{fig:ff-strategy}
\end{figure}

Therefore, we designed a force field tuning process, detailed in Figure \ref{fig:ff-strategy}, which involves the following steps:

\begin{enumerate}
    \item \textbf{Molecule Fragmentation}: Decompose complex organic molecules into fragments containing at least one rotamer using a fragmentation method.
    \item \textbf{Flexible Scan}: Perform flexible scans on the high-accuracy potential surface (QM or NNP) around each rotamer for every fragment. This involves fixing the dihedral angle of the rotatable bond and optimizing bond lengths and angles to obtain relaxed conformations.
    \item \textbf{Parameter Optimization}: Catalog each fragment's fingerprint in a fragment library to identify the dihedral parameters that need optimization. Optimize MM dihedral parameters to minimize the relative error between high-accuracy (QM or NNP) and MM potential surfacesc. The fragments' fingerprints are based on molecular topologies using a node-embedding-based approach, ensuring that similar local environments share identical fingerprints.
    \item \textbf{Parameter Matching}: Match the optimized parameters to complex organic molecules using fingerprints.
\end{enumerate}

\subsection{Step 1. Molecule Fragmentation}
\label{sec:fragmentation}


To facilitate the optimization of molecular force fields, we developed a comprehensive approach for the fragmentation of molecular structures. This process involves several key steps to ensure that the chemical integrity and significant features of the molecules are preserved during fragmentation.



Firstly, a detailed analysis is performed to identify important structural elements such as ring systems and the \emph{ortho} positions relative to specific bonds. This step is crucial for understanding the chemical environment of the molecule and for guiding the subsequent fragmentation process.

To study specific torsion angles, the fragmentation method targets these angles by systematically adding neighboring atoms in layers around the torsion bond. This process encompasses neighboring atoms, functional groups, and connected ring systems, with a keen emphasis on preserving important chemical features such as specific functional groups and ring integrity. Special attention is given to the careful addition of non-carbon and non-hydrogen atoms, carbonyl groups, and specific heteroatoms, ensuring the preservation of chemical significance and functionality in the produced fragments.

Following the determination of the fragments, our method employs RDKit's functionalities to perform the fragmentation, breaking the selected bonds and optionally capping the resulting fragment ends with methyl groups to maintain valency. This ensures that the fragments are chemically meaningful and suitable for subsequent computational analyses.

\begin{figure}
    \centering
    \includegraphics[width=1\linewidth]{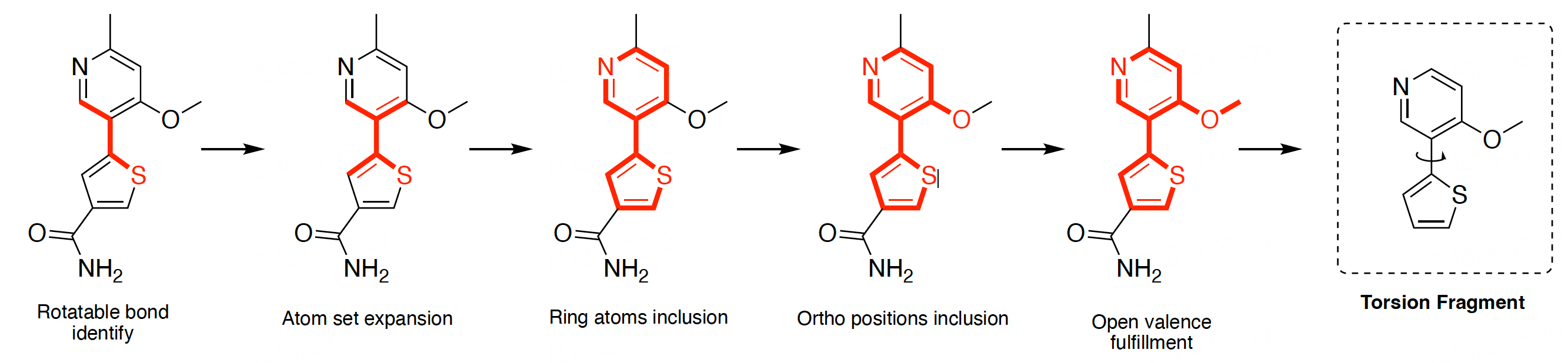}
    \caption{Illustration of fragmentation strategy}
    \label{fig:frag}
\end{figure}

The fragmentation method is illustrated in Figure ~\ref{fig:frag}. This figure shows the step-by-step addition of atoms around a target torsion angle and the resulting fragments that retain important chemical features.

\subsection{Step 2. Torsion Scan with Fine-tuned DPA-2-TB Model}

The most time-consuming part of the outlined process is the flexible scan of the high-accuracy potential surface, which can take days using traditional QM methods. To enhance efficiency, we experimented with constructing specific NNPs for organic small molecules to replicate their potential surfaces. The speed of NNPs far exceeds that of QM methods, allowing for molecular force field fine-tuning to be completed on a laptop within minutes.

There are pioneering works on performing torsion scans of biaryl compounds with NNPs\cite{lahey2020anitorsion}, but their accuracy is not enough for general-purpose torsion scans, and is not easily extensible for building a parameter library.

Large atomic models (LAM) are revolutionizing the way NNPs are produced. The representative LAM, the DPA-2\cite{zhang2023dpa2}, pre-trained on a vast space of chemical structures and \emph{ab initio} labels, can easily generalized to downstream NNP fine-tuning tasks with a data efficiency much higher than other NNPs.

We found that directly fitting the DPA-2 model to QM data of organic molecules will result in a relatively large error, as shown in Figure \ref{fig:dpa2-train}. Inspired by previous works of delta-learning, like OrbNet\cite{qiao2020orbnet} and QRNN-TB\cite{qrnn}, we developed the DPA-2-TB model. These methods utilize semi-empirical methods to describe charge information and long-range interactions that NNPs struggle to learn. In the DPA-2-TB model, delta learning methods are employed to correct GFN2-xTB\cite{gfn2xtb} with the DPA-2 model. 
\begin{align}
     E_i = 
    \mathcal F \Big(  
    \mathcal D_i \big(
    \mathcal R, \mathcal Z
    \big)
    \Big), & \quad \text{original DPA-2}\\
    E_i - E_{i, \mathrm{xTB}} = 
    \mathcal F_{\Delta} \Big(  
    \mathcal D_i \big(
    \mathcal R, \mathcal Z
    \big)
    \Big), & \quad \Delta\text{-learning DPA-2 (i.e. DPA-2-TB)}
\end{align}

DPA-2-TB and DPA-2 models predict the atomic energy contribution based on the atomic numbers $\mathcal Z$ and the coordinates $\mathcal R$. $\mathcal D_i$ represents the descriptor of atom $i$, maps from the atomic numbers and coordinates to a hidden representation that remains invariant under translational, rotational, and permutational (only among atoms with the same atomic number) operations. The structure of the descriptor network and fitting network are detailed in the original reference.
DPA-2-TB and DPA-2 share the same descriptor network. Only the delta-learning fitting network \(F_{\Delta}\) should be trained in the fine-tuning stage. The dataset construction and training details of the DPA-2-TB model are displayed in \nameref{sec:supp}.

In practical use, torsion scans are done by geometry optimizations under a single torsion constraint \(\phi_m = \phi_m^*\), then scan from \( \phi_m^*\in [-180^\circ,+180^\circ]\). We apply torsion scans on DPA-2-TB and original MM potential energy surfaces separately, then calculate the energy differences as training data, to exclude the effects of other degrees of freedom and 1-4 pair interactions.

\begin{equation}
\begin{aligned}
    U_{\mathrm{MM_{new}}}(\phi_m^*) = \min_{\phi_m = \phi_m^*} (E_{\mathrm{DPA-2-TB}} + E_{\mathrm{xTB}}) - \min_{\phi_m = \phi_m^*}E_{\mathrm{MM_{old}}} + U_{\mathrm{MM_{old}}}(\phi_m^*)
\end{aligned}
\end{equation}
Where \(U_{\mathrm{MM}}(\phi_m) = \sum_{n=1}^{6} k_n\cos(n\phi_m + \phi_m^0)\) is the energy contribution of a single torsion.

\subsection{Step 3. Node-Embedding-Based Rotamer Fingerprint for Consistency of Molecular Force Field Parameters}

To maintain robustness and transferability during torsion parameter fitting, structurally similar molecular fragments should share the same parameters as much as possible to minimize potential systematic errors.  

During the development of OpenFF\cite{openff1.0} or other classical forcefields, an SMIRKS-based\cite{smirks,smirnoff} or atom-type-based dihedral parameterization scheme was employed to address this issue, but it required a lot of human effort to design these patterns. We explored a node-embedding-based similarity approach, constructing graphs based on the molecular topologies to assign identical fingerprints to rotamers in the same local environment, thereby achieving parameter consistency and reducing systematic errors.

Given a molecular structure, it is transformed into a graph \(\mathcal{G} = (\mathcal{V}, \mathcal{E})\), where \(\mathcal{V}\) represents the set of vertices (atoms) and \(\mathcal{E}\) the set of edges (bonds). Each vertex \(v_i \in \mathcal{V}\) is associated with a 131-dimensional feature vector \(x_i\), capturing atomic properties including atomic number, valency, and ring membership. The adjacency matrix \(A \in \mathbb{R}^{n \times n}\), with \(n = |\mathcal{V}|\), denotes the connectivity between atoms, where \(A_{ij} = 1\) if a bond exists between atoms \(i\) and \(j\), and \(A_{ij} = 0\) otherwise.

The embedding process employs a randomly generated weight matrix \(W_1 \in \mathbb{R}^{d \times d}\), to map atomic number features into a continuous vector space. The iterative update rule for the embedding of atom \(i\), after \(l\) layers, is defined as:
\[h_i^{(l+1)} = \sigma \left( \sum_{j \in \mathcal{N}(i) \cup \{i\}} A_{ij} \cdot h_j^{(l)}W_1 \right)\]
where \(\mathcal{N}(i)\) denotes the set of neighbors of atom \(i\), \(h_i^{(l)}\) is the feature vector of atom \(i\) at layer \(l\), and \(\sigma\) is a non-linear activation function such as ReLU.

Torsional angles within the molecule are enumerated by analyzing potential rotational bonds. Each torsion is characterized by a tuple \((i, j, k, l)\), representing the indices of the atoms involved. The embedding vector for a torsion is computed as the average of the embeddings of its direct and inverse orientations, \(h_{\mathrm{torsion}} = \frac{1}{2}(h_{ijkl} + h_{lkji})\).

To identify equivalent torsions based on their embeddings, we calculate the Euclidean distance between the embedding vectors of torsion \(m\) and \(n\), \(d_{mn} = ||h_m - h_n||_2\). Torsions are considered equivalent if \(d_{mn} < \epsilon\), where \(\epsilon\) is a predefined threshold.

\subsection{Software Development and Code Availability}

To evaluate the NNP's effectiveness, we developed a stable and reproducible testing process in NNP benchmark tool (\href{https://github.com/WangXinyan940/schrodinger-nnp-benchmark}{https://github.com/WangXinyan940/schrodinger-nnp-benchmark}). Users input an ASE calculator, allowing us to assess the calculator's accuracy based on Schodinger's QRNN work, comparing it with QM results, semi-empirical methods, and NNPs published by Schrodinger. 

Subsequently, we implemented and open-sourced the dihedral parameter fine-tuning tool, DeePDih (\href{https://github.com/WangXinyan940/DeePDih}{https://github.com/deepmodeling/DeePDih)}, based on the dihedral optimization process. It allows for structure optimization using user-defined ASE calculators, and saves optimized dihedral parameters in a library, enabling parameter lookup and replacement for new molecules. Its structure optimization tool is JIT accelerated, achieving more than a fivefold performance increase.

\section{Results and Discussions}

\subsection{Evaluation of DPA-2-TB Model}

\begin{figure}
    \centering
    \includegraphics[width=1\linewidth]{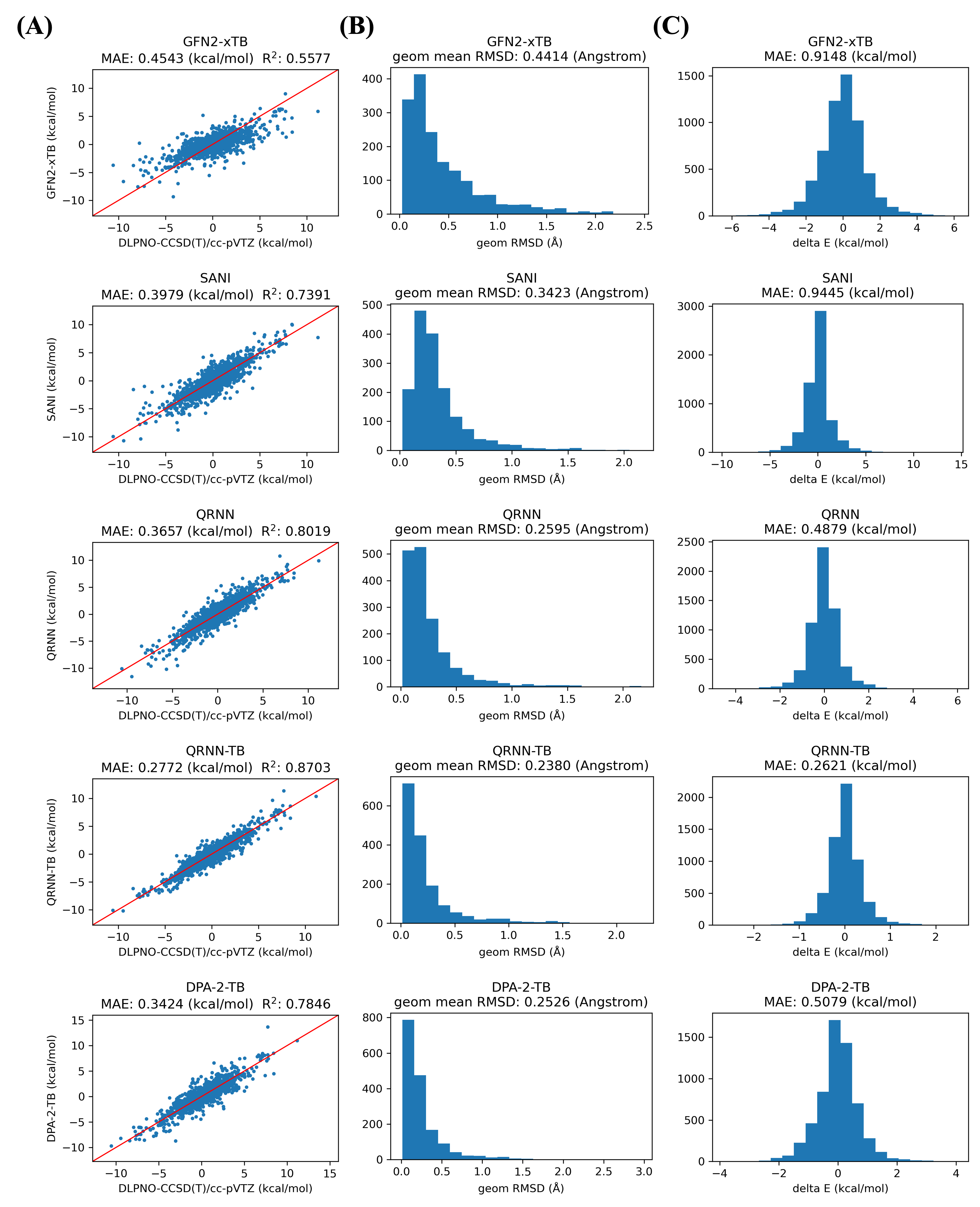}
    \caption{Benchmark of GFN2-xTB, Schrodinger-ANI, QRNN, QRNN-TB, and DPA-2-TB on three datasets: (A) conformational energy test on the Hutchison dataset, (B) geometry optimization test, and (C) dihedral scan test. DPA-2-TB significantly outperforms xTB and ANI, achieves the same accuracy as QRNN, but is slightly inferior to QRNN-TB.}
    \label{fig:dpa2-eval}
\end{figure}

\subsubsection{Accuracy of Relative Conformational Energies}

We proceed to evaluate the energy prediction capabilities of models. Our objective is to confirm the models' efficacy in accurately representing the geometries and energies of organic molecule conformers and their ionic counterparts. Our primary focus lies in assessing model transferability by testing on molecules not included in the training set, for the key application of calibrating torsion parameters in conventional force fields. 

To this end, we primarily focus on exploring the relative conformational energies of flexible molecules. Instead of examining the heights of torsion barriers as torsion scans do, relative conformational energies delve into the potential energy surface's minima. To assess the conformational accuracy of our models, we utilize a test set curated by Folmsbee and Hutchison\cite{hutchison2020} then recalculated by Schrodinger\cite{qrnn}, comprising 576 neutral and 81 charged molecules, all anticipated to exhibit conformational variability. Hutchison et al. conducted a conformational search for these molecules, identifying up to ten low-energy conformers for each.

The test set includes molecules with up to 50 heavy atoms and 23 rotatable bonds, posing a significant challenge to our models' generalizability, given the smaller size of our training data. We calculate the Mean Absolute Error (MAE) and the square of the Pearson correlation coefficient (R²) across the conformer sets. We employ the median values of these metrics to evaluate our ability to replicate reference energies and the accuracy of conformation ranking. Hutchison's findings indicate that DFT methods often achieve median R² values above 0.8 and MAE below 0.3 kcal/mol; and as Figure \ref{fig:dpa2-eval}(A) shows, the DPA-2-TB model and QRNN-TB model achieve an MAE of 0.2775 kcal/mol and 0.3555 kcal/mol, outperforming the best empirical approaches (GFN2-xTB) and Schrodinger-ANI model.

\subsubsection{Accuracy of Optimized Geometries}

To evaluate the precision of optimized geometries, we adopted the dataset from QRNN comprising ionic conformers of drug-like molecules. These conformers underwent geometry optimization to evaluate the accuracy of their optimized geometries against \(\omega\)B97X-D/6-31G* benchmarks, measuring discrepancies in bond lengths, angles, torsions, and RMSDs across all Cartesian coordinates, as well as in relative energies and median R² values.

Our findings, detailed in Figure \ref{fig:dpa2-eval}(B), indicate improvements in errors across bond lengths, angles, torsions, and Cartesian RMSDs from semiempirical methods to DPA-2-TB, achieving remarkably low errors. The results underscore the exceptional congruence of the delta-learned models' geometries with DFT references, advocating for their potential in future research on conformer energy ranking.

\subsubsection{Accuracy of Relative Torsion Energies}

We have conducted accuracy assessments on new torsion scans for molecules with drug-like properties selected from QRNN's benchmark set\cite{qrnn}, consisting of 388 distinct relaxed torsion scans (optimized using DFT at the reference level) of "charged" species, and 112 distinct scans of "neutral" molecules.

As shown in Figure \ref{fig:dpa2-eval}(C), the DPA-2-TB model has reached an energy MAE of 0.3424 kcal/mol, outperforming that of empirical models like GFN2-xTB\cite{gfn2xtb}, and Schrodinger-ANI\cite{schrodinger-ani,lahey2020anitorsion} model, though not better than the delta learning QRNN-TB model.

\subsection{FEP Calculation Results}


We demonstrate the effects of the DPA-2-TB corrected force field on free energy perturbation (FEP) calculations with a couple of examples\cite{wang2015accurate}. We perform the MD simulations with a modified version of Gromacs 2021.2\cite{gromacs}. The AMBER99SB-ILDN\cite{99sb-ildn} and general AMBER force field (GAFF) v2 force fields\cite{gaff} are adopted. Both protein-ligand complex and free ligand systems are solvated in an orthogonal box with a water buffer size of 1 nm. The cutoffs for both VDW and Coulomb interactions are 1.1 nm. PME is used to calculate long-range electrostatics. The softcore is applied on both Coulomb and Vdw interactions. Systems are just neutralized by adding Na¸ and Cl ions.

16 replicas with adaptive mesh are set. Bond, vdw, charge are coupled with \(\lambda\) and uniformly transform from A to B. In each replica, a energy minimization (EM) is first performed until the maximal force is less than 100 kJ/mol/nm. Next, a 20 ps NVT and 20 ps NPT are included for equilibration. Finally, a 2 fs time step MD simulation lasts for 5 ns, and the MBAR\cite{shirts2008mbar} method is utilized to obtain the free energy difference.

\begin{figure}
    \centering
    \includegraphics[width=1.0\linewidth]{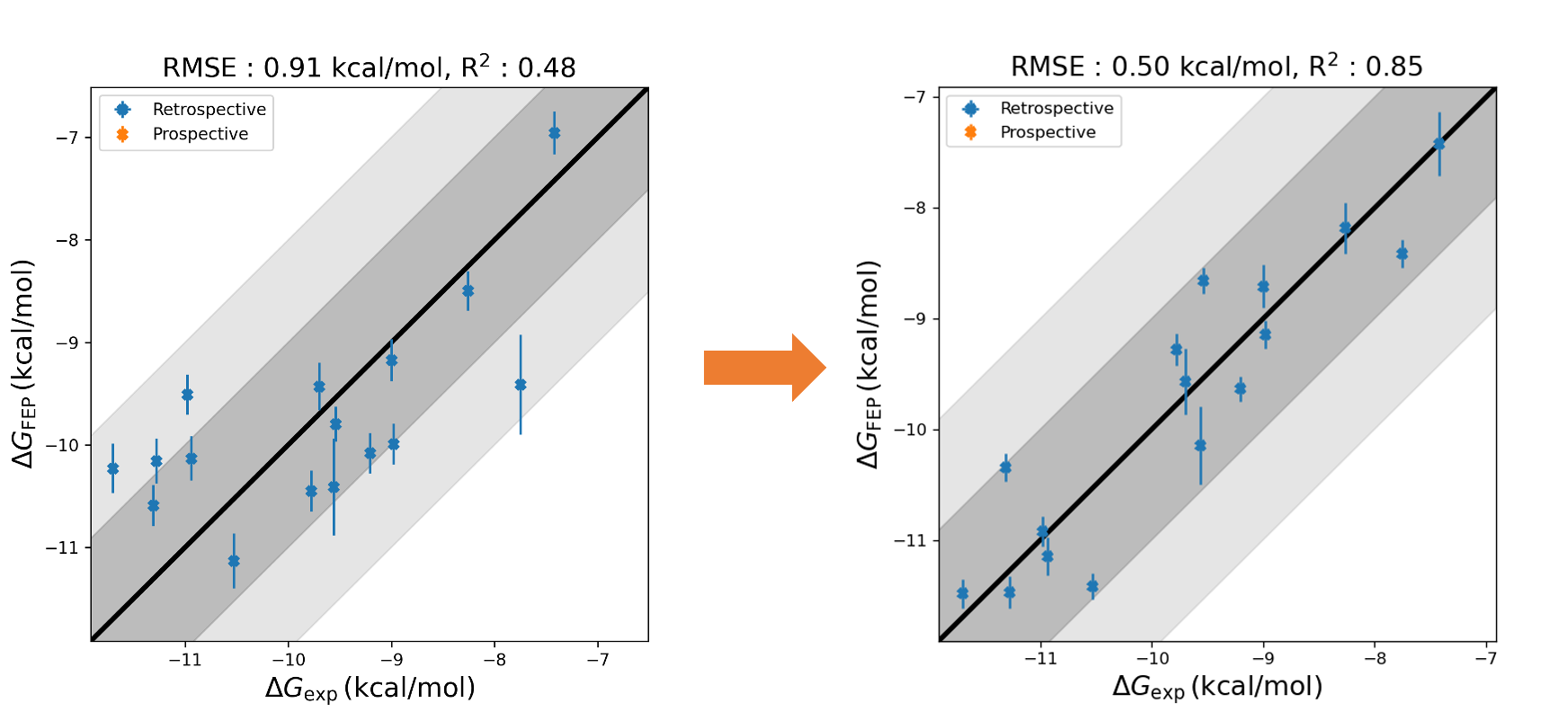}
    \caption{Utilizing the (left) original and the (right) DPA-2-TB fine-tuned GAFF2 force fields, the relative free energies of the TYK2 test system were calculated. Following the fine-tuning, there was a notable enhancement in the precision of free energy predictions for various drug molecules.
}
    \label{fig:fep-tyk2}
\end{figure}

As shown in Figure \ref{fig:fep-tyk2}, the RMSE over the Tyk2 inhibitor series improves from 0.91 kcal/mol with GAFF2 to 0.50 kcal/mol with DPA-2-TB corrected GAFF2. 

\begin{figure}
    \centering
    \includegraphics[width=1\linewidth]{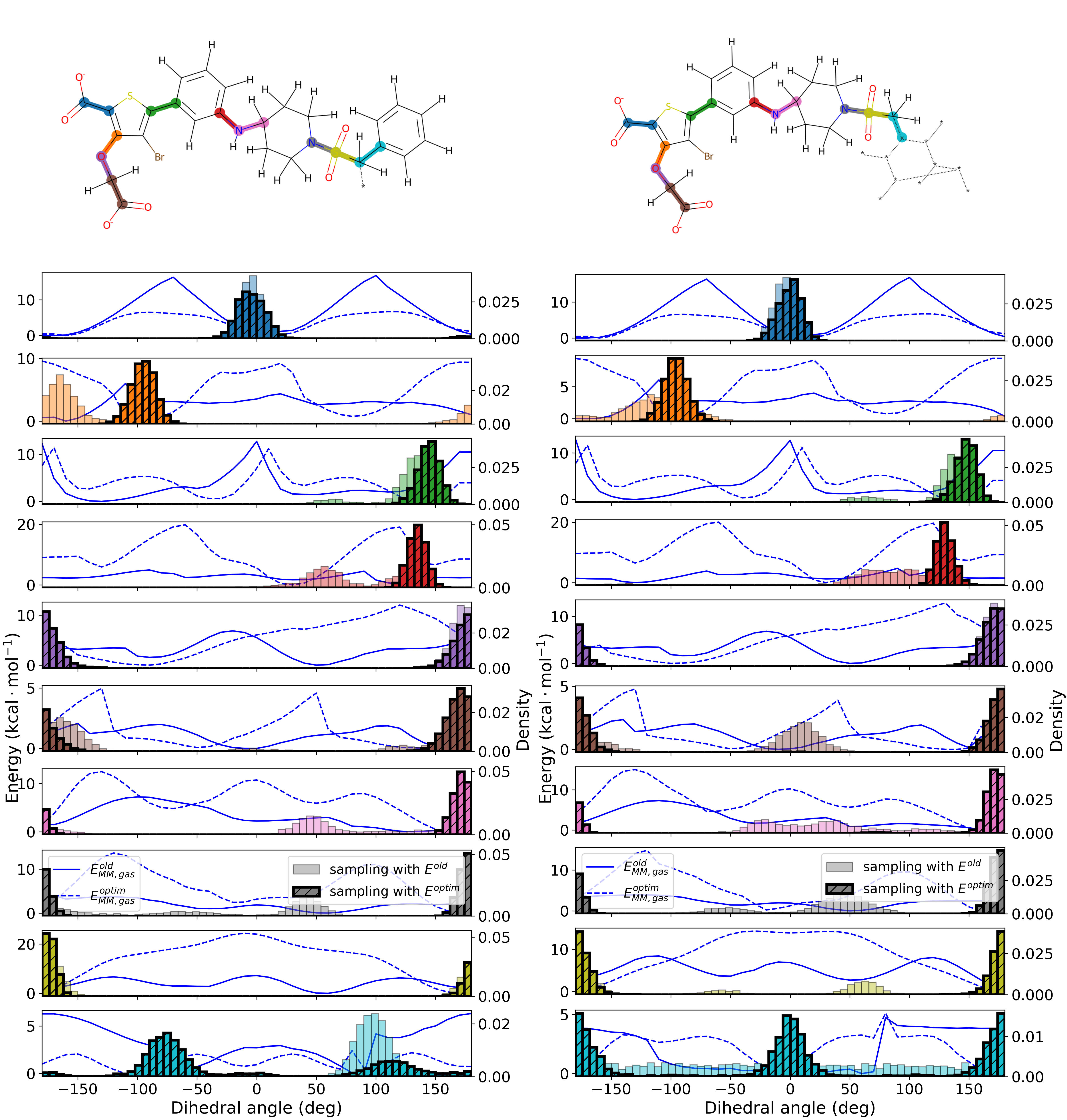}
    \caption{The dihedral angle sampling in protein-ligand complex of molecule \textbf{20667} (left) and \textbf{23482} (right), with GAFF2 (\(E^{\mathrm{old}}\)) and DPA-2-TB corrected GAFF2 (\(E^{\mathrm{optim}}\)). The colored bars show the distribution in MD trajectories of the highlighted torsion, the blue lines are MM torsion energy profiles in gas phase. 
    The grey one indicates the sulfonamide group's optimal conformer has been corrected from \emph{gauche} to \emph{trans}.}
    \label{fig:lig_torsion_ff}
\end{figure}

\begin{figure}
    \centering
    \includegraphics[width=0.48\linewidth]{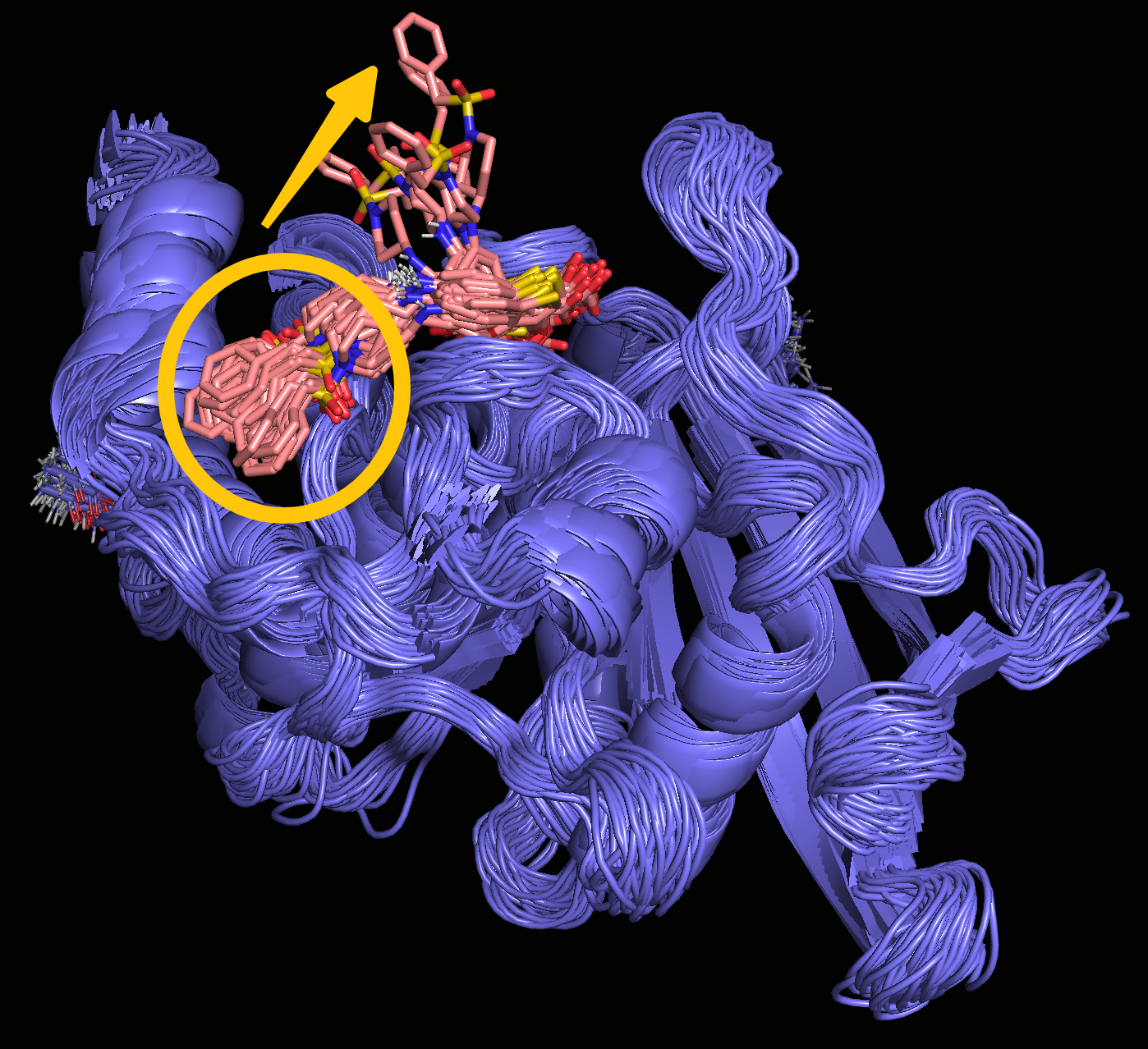}
    \includegraphics[width=0.48\linewidth]{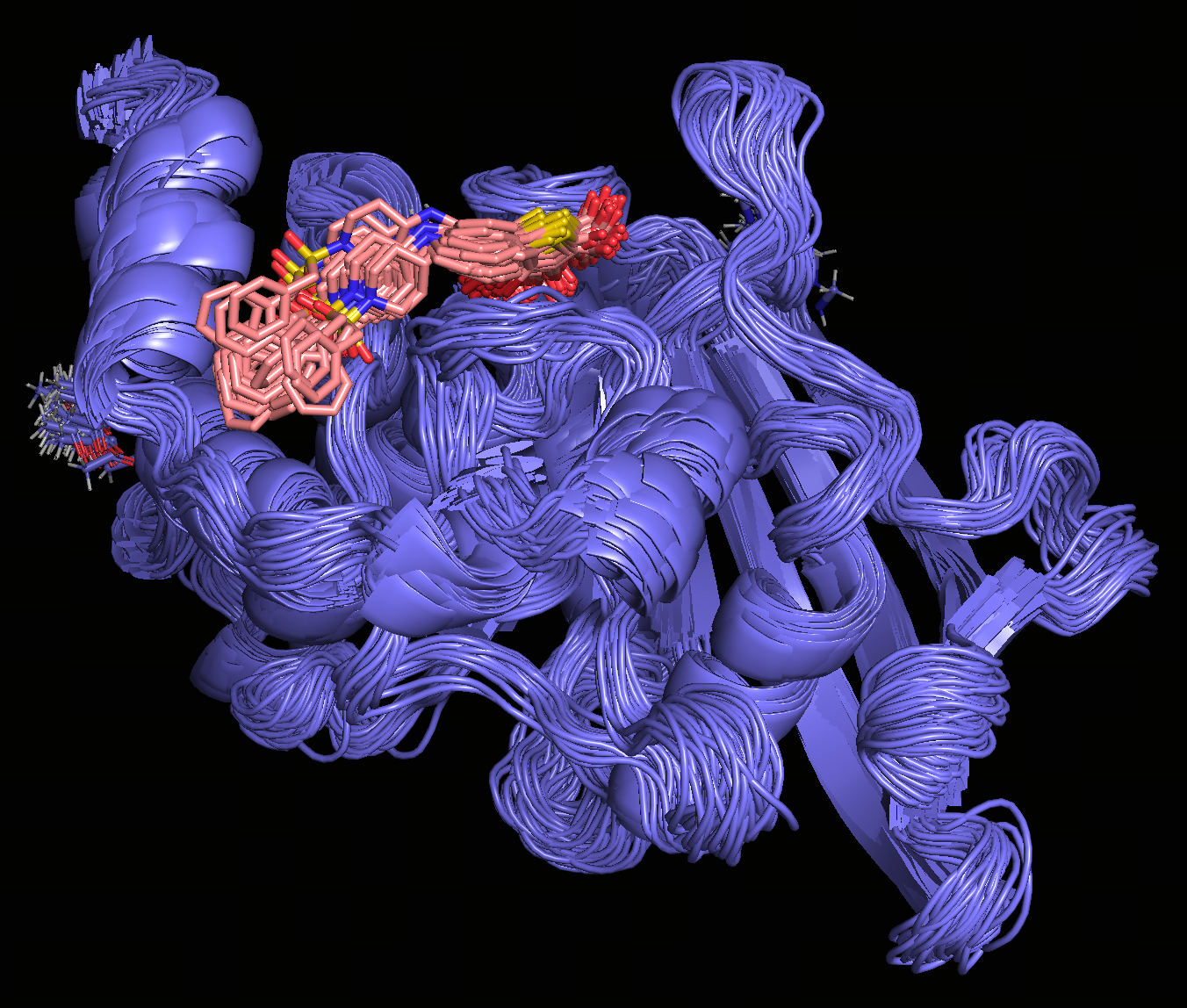}
    \caption{Sampling protein-ligand complex of PTP1B and molecule \textbf{20667} with GAFF2 (left) and DPA-2-TB corrected GAFF2 (right). The optimized force field keeps the ligand stable in the subpocket.}
    \label{fig:pocket}
\end{figure}

A substantial improvement is also seen in PTP1B, where the RMSE is 1.15 kcal/mol in GAFF2 and 0.91 kcal/mol in DPA-2-TB corrected GAFF2. 
\begin{figure}
    \centering
    \includegraphics[width=1\linewidth]{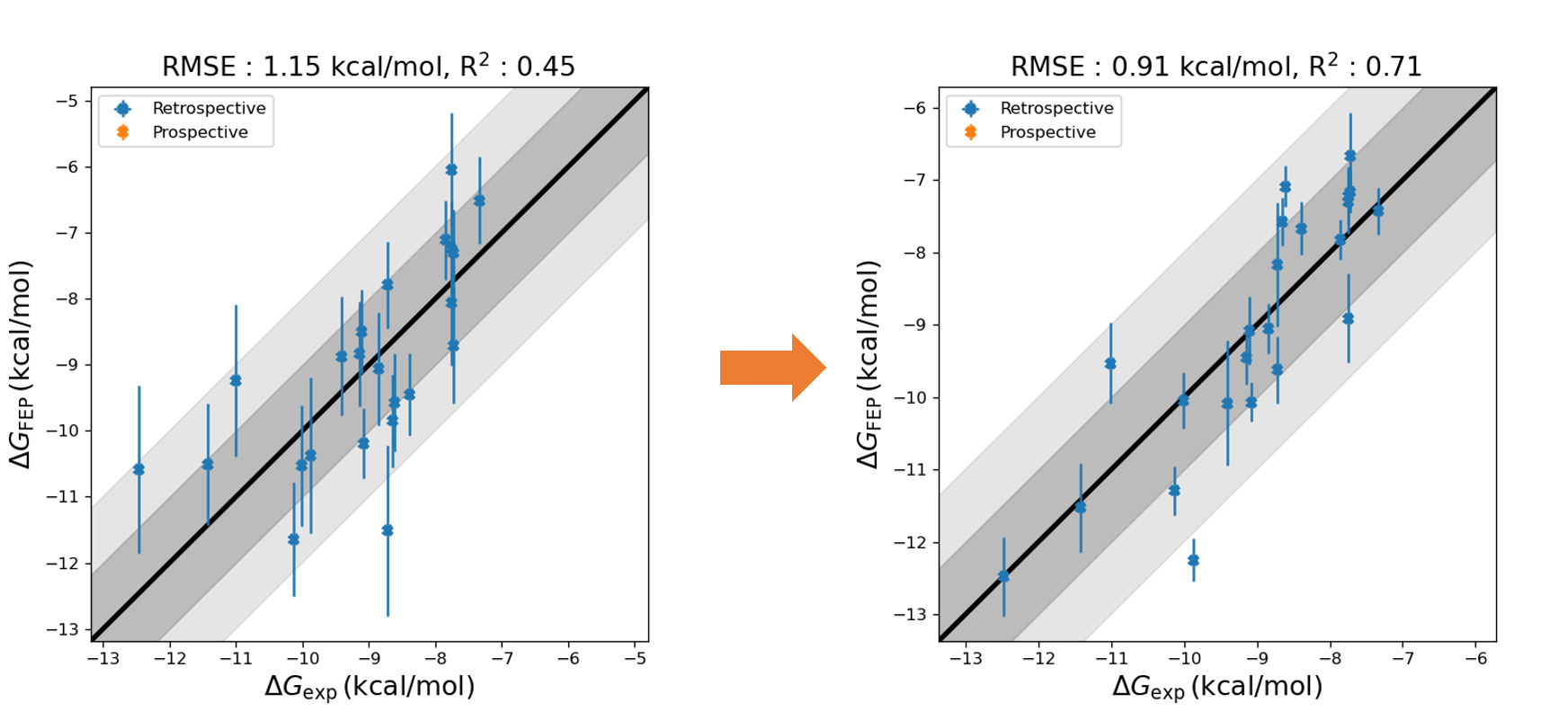}
    \caption{Utilizing the (left) original and the (right) DPA-2-TB fine-tuned GAFF2 force fields, the relative free energies of the PTP1B test system were calculated. Following the fine-tuning, there was a notable enhancement in the precision of free energy predictions for various drug molecules.}
    \label{fig:fep-ptp1b}
\end{figure}
More specifically, in a representative edge from molecule \textbf{20667} to \textbf{23482}, as illustrated in Figure \ref{fig:lig_torsion_ff}, the absolute difference of FEP calculated \(\Delta\Delta G\) compared to experimental \(\Delta\Delta G\), has lowered from 2.74 kcal/mol with GAFF2 to 0.40 kcal/mol with DPA-2-TB corrected GAFF2. The MM torsion profiles show that the conformation distribution of the sulfonamide group (highlighted gray) in the protein-ligand complex has been corrected from \emph{gauche} to \emph{trans}, avoiding the dissociation of the phenyl group to the subpocket, as shown in Figure \ref{fig:pocket}, thus keeping the conformational sampling in a reasonable region.



\section{Conclusion}

In this study, we developed a novel approach for optimizing molecular force fields by combining the fine-tuned DPA-2 NNP with a node-embedding-based similarity metric. 
Traditional force fields often struggle with accurately representing novel molecules and complex chemical environments due to the labor-intensive process of manually setting optimization parameters and the high computational cost of QM calculations. 
Our method addresses these challenges by optimizing force field parameters on-the-fly, ensuring uniform parameter consistency across various molecular fragments, and significantly reducing computational costs.

Our approach involves several key steps to enhance both precision and efficiency. 
First, we fine-tuned the DPA-2 pre-trained model to predict high-accuracy potential surfaces. This model leverages delta learning techniques to correct semi-empirical QM calculations, allowing it to achieve precision comparable to QM methods but at a fraction of the computational cost. 
Second, we developed a node-embedding-based similarity metric to replace the traditional, labor-intensive process of manually defining atom types or substructure types. By transforming molecular structures into graph representations, where atoms and bonds are treated as nodes and edges, respectively, our method ensures that structurally similar molecular fragments share identical parameters. 
This approach enhances parameter consistency across different molecules and reduces systematic errors.

To demonstrate the effectiveness of our method, we applied it to the TYK2 inhibitor and PTP1B systems. 
The results showed significant improvements in FEP calculations, with the RMSE reduced from 0.91 kcal/mol to 0.50 kcal/mol for the TYK2 inhibitor series and from 1.15 kcal/mol to 0.91 kcal/mol for the PTP1B system. 
These findings highlight the potential of our approach to enhance the precision of molecular simulations, providing more accurate energy and free energy predictions compared to existing models.

Looking forward, we plan to extend this work in several directions. 
First, we aim to scan as many drug-like molecules as possible using our method to develop a relatively universal intramolecular force field. This effort will provide a comprehensive and broadly applicable tool for the computational chemistry community. 
Second, we intend to generalize this method to grid-based energy correction maps (CMAP)\cite{cmap1,cmap2}, which will make it possible to calculate the PES of ring flipping and dihedral angle correlations in biomacromolecules accurately.
Third, we plan to conduct an independent study focused on improving intermolecular interactions, addressing specific challenges such as halogen bonds and heterocycles in drug molecules, and exploring new schemes for applying optimized force fields in various chemical scenarios.





\begin{suppinfo}
\label{sec:supp}

\subsection{Fitting in Rotamer Space to Avoid Inconsistencies Between High-Accuracy Potential Surfaces and Molecular Force Fields}
\label{sec:inconsistency}

Fig. \ref{fig:qm_mm_diff} shows a comparative analysis of the energy differences on MM potential surfaces between QM-optimized structures and MM-optimized structures. The study highlights the potential risks of additional energy uncertainties (about 2 kcal/mol) when directly using QM-optimized conformations. 

\begin{figure}
    \centering
    \includegraphics[width=1.0\linewidth]{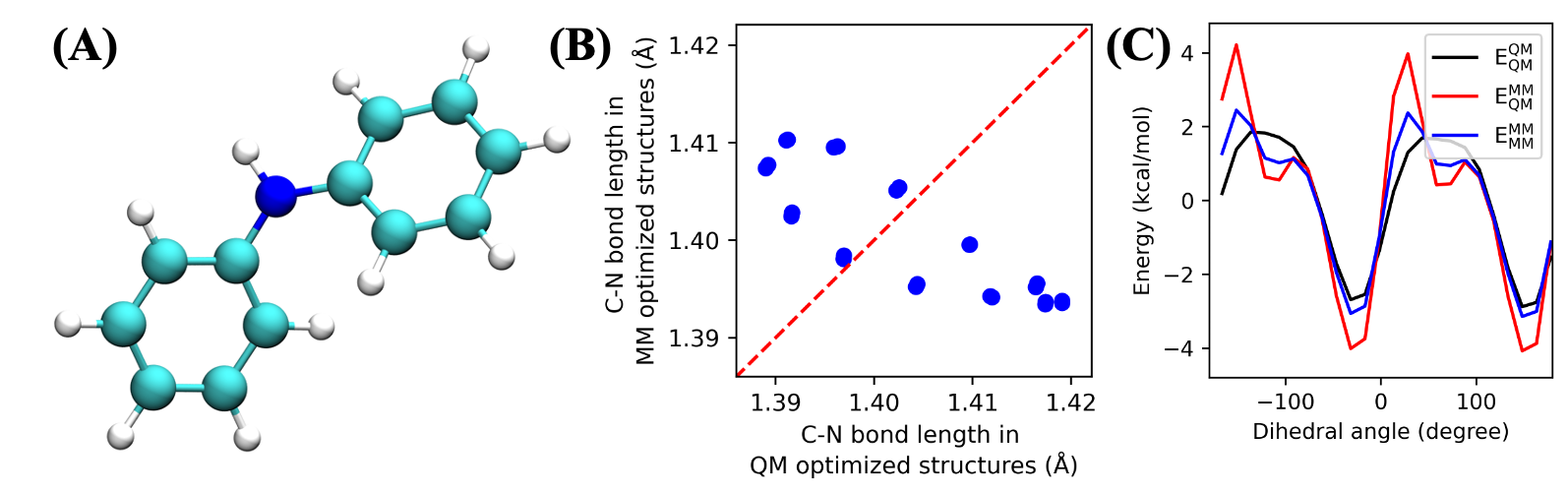}
    
    \caption{(A) The molecule showing the differences between QM and MM potential energy surfaces (PES) is subjected to a dihedral angle scan along the C-N bond. The QM PES is computed using the \(\omega\)B97x-d/def2-SVP method, while the MM PES is obtained using the GAFF2 force field.
(B) The variation in C-N bond lengths among the conformations from the dihedral scan on both QM and MM PES is investigated. The trend in C-N bond lengths in the relaxed structures on the QM and MM PES is found to be inconsistent.
(C) The energy profiles from the dihedral scan are obtained using the QM method on the QM-optimized conformations (E$^{\text{QM}}_{\text{QM}}$), the MM method on the QM-optimized conformations (E$^{\text{MM}}_{\text{QM}}$), and the MM method on the MM-optimized conformations (E$^{\text{MM}}_{\text{MM}}$). The E$^{\text{MM}}_{\text{MM}}$ energy profile exhibits smaller energy deviations compared to the E$^{\text{MM}}_{\text{QM}}$ energy profile.}
    \label{fig:qm_mm_diff}
\end{figure}

Meanwhile, the critical role of rotatable bonds in the conformational changes of organic molecules is emphasized, with a detailed analysis using Alanine as an example (Fig. \ref{fig:ala2-pes}). High-dimensional organic molecules can be effectively described in low-dimensional spaces through rotamers, ensuring consistency between traditional force fields and high-accuracy methods in rotamer space.

\begin{figure}
    \centering
    \includegraphics[width=0.6\linewidth]{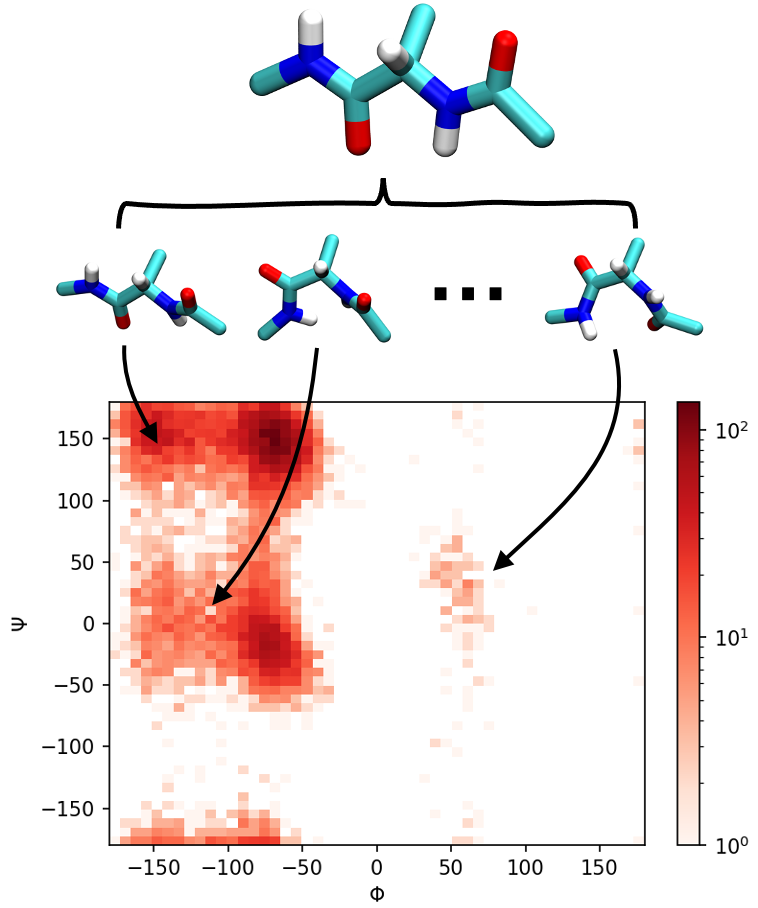}
    \caption{The high-dimensional conformational space of the Alanine dipeptide can be effectively characterized by the two-dimensional Ramachandran plot.}
    \label{fig:ala2-pes}
\end{figure}

\subsection{Data Set Construction}
\label{sec:dataset}

To construct an NNP suitable for organic small molecules, we specially designed the dataset for this NNP, including the coverage range of the pre-training dataset and model optimization based on a druglike dataset. 

The SPICE dataset\cite{spice}, based on published structures of biomolecules, drug molecules, and their complexes generated with classical force fields and MD, serves as an ideal pre-training dataset, increasing the chemical space coverage of the pre-trained model. We selected a representative subset of conformations (approximately 80\% of the total dataset) that are consistent with chemical intuition for pre-training. Energies and forces are labeled at \(\omega\)B97M-D3BJ/def2-TZVPPD\cite{wb97m-v,rappoport2010def2basis} level of theory.

The fine-tuning process utilized a druglike dataset, a molecular fragment database built on the ChEMBL dataset\cite{chembl2012} with 600,000 molecular conformations. All fragments are generated by the method described in section \nameref{sec:fragmentation}. This dataset is generated by DP-GEN\cite{zhang2020dpgen}'s active learning strategy for more continuity in conformation space, making it suitable for fine-tuning potential surfaces. Energies and forces are labeled at \(\omega\)B97X-D/def2-SVP\cite{wb97x-D,rappoport2010def2basis} level of theory. All datasets are publicly available on the OpenLAM project.

\subsection{Training and Evaluation of DPA-2-TB Model}

The pre-trained DPA-2 model is adopted from the 2024Q1 version of the OpenLAM project, trained on a broad spectrum of chemical and configurational datasets\cite{zhang2023dpa2} including the processed SPICE dataset described in section \nameref{sec:dataset}. 

In the fine-tuning stage, the root mean squared error (RMSE) of energies and forces during fine-tuning is shown in Figure \ref{fig:dpa2-train}(A) and Figure \ref{fig:dpa2-train}(B). DPA-2-TB fine-tuning with the difference between DFT and xTB, can reach an energy RMSE of 0.04 kcal/mol/atom, and force RMSE of 0.8 kcal/mol/Angstrom. Direct DPA2 fine-tuning with DFT energies and forces can only reach an RMSE of 0.1 kcal/mol/atom and 2.0 kcal/mol/Angstrom.

\begin{figure}
    \centering
    \includegraphics[width=1.00\linewidth]{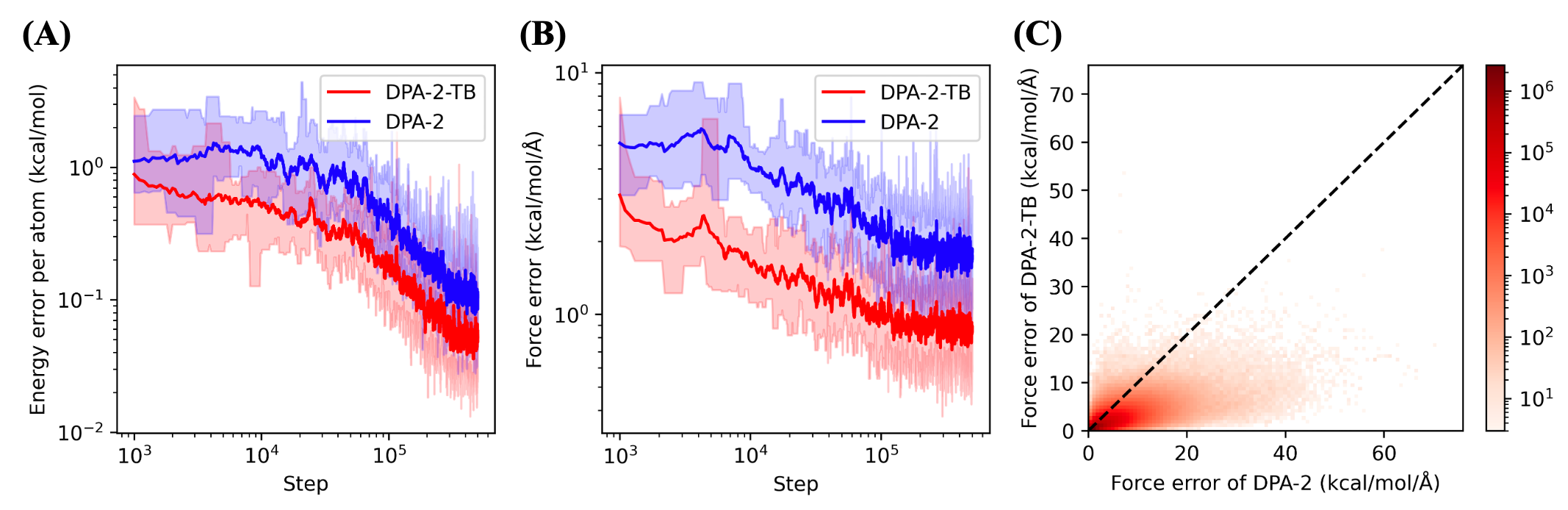}
    \caption{The loss of (A) energy and (B) forces when fine-tuning DPA2 and DPA-2-TB. (C) The force prediction error of DPA2 and DPA-2-TB. The DPA-2-TB model generally exhibits higher accuracy compared to the original DPA2 model.}
    \label{fig:dpa2-train}
\end{figure}

After fine-tuning, the forces predicted by the DPA-2-TB model are much more consistent with DFT calculations, shown in Figure \ref{fig:dpa2-train}(C).

\end{suppinfo}


\begin{mcitethebibliography}{43}
\providecommand*\natexlab[1]{#1}
\providecommand*\mciteSetBstSublistMode[1]{}
\providecommand*\mciteSetBstMaxWidthForm[2]{}
\providecommand*\mciteBstWouldAddEndPuncttrue
  {\def\EndOfBibitem{\unskip.}}
\providecommand*\mciteBstWouldAddEndPunctfalse
  {\let\EndOfBibitem\relax}
\providecommand*\mciteSetBstMidEndSepPunct[3]{}
\providecommand*\mciteSetBstSublistLabelBeginEnd[3]{}
\providecommand*\EndOfBibitem{}
\mciteSetBstSublistMode{f}
\mciteSetBstMaxWidthForm{subitem}{(\alph{mcitesubitemcount})}
\mciteSetBstSublistLabelBeginEnd
  {\mcitemaxwidthsubitemform\space}
  {\relax}
  {\relax}

\bibitem[Zhang \latin{et~al.}(2018)Zhang, Han, Wang, Car, and E]{deepmd-prl}
Zhang,~L.; Han,~J.; Wang,~H.; Car,~R.; E,~W. Deep Potential Molecular Dynamics: A Scalable Model with the Accuracy of Quantum Mechanics. \emph{Physical Review Letters} \textbf{2018}, \emph{120}, 143001\relax
\mciteBstWouldAddEndPuncttrue
\mciteSetBstMidEndSepPunct{\mcitedefaultmidpunct}
{\mcitedefaultendpunct}{\mcitedefaultseppunct}\relax
\EndOfBibitem
\bibitem[Schütt \latin{et~al.}(2017)Schütt, Kindermans, Sauceda, Chmiela, Tkatchenko, and Müller]{schütt2017schnet}
Schütt,~K.~T.; Kindermans,~P.-J.; Sauceda,~H.~E.; Chmiela,~S.; Tkatchenko,~A.; Müller,~K.-R. SchNet: A continuous-filter convolutional neural network for modeling quantum interactions. 2017\relax
\mciteBstWouldAddEndPuncttrue
\mciteSetBstMidEndSepPunct{\mcitedefaultmidpunct}
{\mcitedefaultendpunct}{\mcitedefaultseppunct}\relax
\EndOfBibitem
\bibitem[Devereux \latin{et~al.}(2020)Devereux, Smith, Huddleston, Barros, Zubatyuk, Isayev, and Roitberg]{ani2x}
Devereux,~C.; Smith,~J.~S.; Huddleston,~K.~K.; Barros,~K.; Zubatyuk,~R.; Isayev,~O.; Roitberg,~A.~E. Extending the Applicability of the ANI Deep Learning Molecular Potential to Sulfur and Halogens. \emph{Journal of Chemical Theory and Computation} \textbf{2020}, \emph{16}, 4192--4202\relax
\mciteBstWouldAddEndPuncttrue
\mciteSetBstMidEndSepPunct{\mcitedefaultmidpunct}
{\mcitedefaultendpunct}{\mcitedefaultseppunct}\relax
\EndOfBibitem
\bibitem[Qiao \latin{et~al.}(2020)Qiao, Welborn, Anandkumar, Manby, and Miller]{qiao2020orbnet}
Qiao,~Z.; Welborn,~M.; Anandkumar,~A.; Manby,~F.~R.; Miller,~r.,~T.~F. OrbNet: Deep learning for quantum chemistry using symmetry-adapted atomic-orbital features. \emph{J Chem Phys} \textbf{2020}, \emph{153}, 124111\relax
\mciteBstWouldAddEndPuncttrue
\mciteSetBstMidEndSepPunct{\mcitedefaultmidpunct}
{\mcitedefaultendpunct}{\mcitedefaultseppunct}\relax
\EndOfBibitem
\bibitem[Batzner \latin{et~al.}(2022)Batzner, Musaelian, Sun, Geiger, Mailoa, Kornbluth, Molinari, Smidt, and Kozinsky]{nequip}
Batzner,~S.; Musaelian,~A.; Sun,~L.; Geiger,~M.; Mailoa,~J.~P.; Kornbluth,~M.; Molinari,~N.; Smidt,~T.~E.; Kozinsky,~B. E(3)-equivariant graph neural networks for data-efficient and accurate interatomic potentials. \emph{Nature communications} \textbf{2022}, \emph{13}, 2453--NA\relax
\mciteBstWouldAddEndPuncttrue
\mciteSetBstMidEndSepPunct{\mcitedefaultmidpunct}
{\mcitedefaultendpunct}{\mcitedefaultseppunct}\relax
\EndOfBibitem
\bibitem[Kovács \latin{et~al.}(2023)Kovács, Moore, Browning, Batatia, Horton, Kapil, Witt, Magdău, Cole, and Csányi]{kovács2023maceoff23}
Kovács,~D.~P.; Moore,~J.~H.; Browning,~N.~J.; Batatia,~I.; Horton,~J.~T.; Kapil,~V.; Witt,~W.~C.; Magdău,~I.-B.; Cole,~D.~J.; Csányi,~G. MACE-OFF23: Transferable Machine Learning Force Fields for Organic Molecules. 2023\relax
\mciteBstWouldAddEndPuncttrue
\mciteSetBstMidEndSepPunct{\mcitedefaultmidpunct}
{\mcitedefaultendpunct}{\mcitedefaultseppunct}\relax
\EndOfBibitem
\bibitem[Lahey and Rowley(2020)Lahey, and Rowley]{2020nnpmm}
Lahey,~S.~J.; Rowley,~C.~N. Simulating protein-ligand binding with neural network potentials. \emph{Chem Sci} \textbf{2020}, \emph{11}, 2362--2368\relax
\mciteBstWouldAddEndPuncttrue
\mciteSetBstMidEndSepPunct{\mcitedefaultmidpunct}
{\mcitedefaultendpunct}{\mcitedefaultseppunct}\relax
\EndOfBibitem
\bibitem[Sabanes~Zariquiey \latin{et~al.}(2024)Sabanes~Zariquiey, Galvelis, Gallicchio, Chodera, Markland, and De~Fabritiis]{chodera2024nnpfep}
Sabanes~Zariquiey,~F.; Galvelis,~R.; Gallicchio,~E.; Chodera,~J.~D.; Markland,~T.~E.; De~Fabritiis,~G. Enhancing Protein-Ligand Binding Affinity Predictions Using Neural Network Potentials. \emph{J Chem Inf Model} \textbf{2024}, \emph{64}, 1481--1485\relax
\mciteBstWouldAddEndPuncttrue
\mciteSetBstMidEndSepPunct{\mcitedefaultmidpunct}
{\mcitedefaultendpunct}{\mcitedefaultseppunct}\relax
\EndOfBibitem
\bibitem[Wang \latin{et~al.}(2004)Wang, Wolf, Caldwell, Kollman, and Case]{gaff}
Wang,~J.; Wolf,~R.~M.; Caldwell,~J.~W.; Kollman,~P.~A.; Case,~D.~A. Development and testing of a general amber force field. \emph{Journal of computational chemistry} \textbf{2004}, \emph{25}, 1157--1174\relax
\mciteBstWouldAddEndPuncttrue
\mciteSetBstMidEndSepPunct{\mcitedefaultmidpunct}
{\mcitedefaultendpunct}{\mcitedefaultseppunct}\relax
\EndOfBibitem
\bibitem[He \latin{et~al.}(2020)He, Man, Yang, Lee, and Wang]{gaff-abcg2}
He,~X.; Man,~V.~H.; Yang,~W.; Lee,~T.-S.; Wang,~J. A fast and high-quality charge model for the next generation general AMBER force field. \emph{The Journal of chemical physics} \textbf{2020}, \emph{153}, 114502--NA\relax
\mciteBstWouldAddEndPuncttrue
\mciteSetBstMidEndSepPunct{\mcitedefaultmidpunct}
{\mcitedefaultendpunct}{\mcitedefaultseppunct}\relax
\EndOfBibitem
\bibitem[Vanommeslaeghe and MacKerell(2012)Vanommeslaeghe, and MacKerell]{cgenff}
Vanommeslaeghe,~K.; MacKerell,~A.~D. Automation of the CHARMM General Force Field (CGenFF) I: bond perception and atom typing. \emph{Journal of chemical information and modeling} \textbf{2012}, \emph{52}, 3144--3154\relax
\mciteBstWouldAddEndPuncttrue
\mciteSetBstMidEndSepPunct{\mcitedefaultmidpunct}
{\mcitedefaultendpunct}{\mcitedefaultseppunct}\relax
\EndOfBibitem
\bibitem[Soteras~Gutierrez \latin{et~al.}(2016)Soteras~Gutierrez, Lin, Vanommeslaeghe, Lemkul, Armacost, Brooks, and MacKerell]{cgenff2016halogen}
Soteras~Gutierrez,~I.; Lin,~F.~Y.; Vanommeslaeghe,~K.; Lemkul,~J.~A.; Armacost,~K.~A.; Brooks,~r.,~C.~L.; MacKerell,~J.,~A.~D. Parametrization of halogen bonds in the CHARMM general force field: Improved treatment of ligand-protein interactions. \emph{Bioorg Med Chem} \textbf{2016}, \emph{24}, 4812--4825\relax
\mciteBstWouldAddEndPuncttrue
\mciteSetBstMidEndSepPunct{\mcitedefaultmidpunct}
{\mcitedefaultendpunct}{\mcitedefaultseppunct}\relax
\EndOfBibitem
\bibitem[Qiu \latin{et~al.}(2021)Qiu, Smith, Boothroyd, Jang, Hahn, Wagner, Bannan, Gokey, Lim, Stern, Rizzi, Tjanaka, Tresadern, Lucas, Shirts, Gilson, Chodera, Bayly, Mobley, and Wang]{openff1.0}
Qiu,~Y. \latin{et~al.}  Development and Benchmarking of Open Force Field v1.0.0-the Parsley Small-Molecule Force Field. \emph{Journal of chemical theory and computation} \textbf{2021}, \emph{17}, 6262--6280\relax
\mciteBstWouldAddEndPuncttrue
\mciteSetBstMidEndSepPunct{\mcitedefaultmidpunct}
{\mcitedefaultendpunct}{\mcitedefaultseppunct}\relax
\EndOfBibitem
\bibitem[Boothroyd \latin{et~al.}(2023)Boothroyd, Behara, Madin, Hahn, Jang, Gapsys, Wagner, Horton, Dotson, Thompson, Maat, Gokey, Wang, Cole, Gilson, Chodera, Bayly, Shirts, and Mobley]{openff2.0}
Boothroyd,~S. \latin{et~al.}  Development and Benchmarking of Open Force Field 2.0.0: The Sage Small Molecule Force Field. \emph{J Chem Theory Comput} \textbf{2023}, \emph{19}, 3251--3275\relax
\mciteBstWouldAddEndPuncttrue
\mciteSetBstMidEndSepPunct{\mcitedefaultmidpunct}
{\mcitedefaultendpunct}{\mcitedefaultseppunct}\relax
\EndOfBibitem
\bibitem[Jorgensen \latin{et~al.}(1996)Jorgensen, Maxwell, and Tirado-Rives]{opls-aa}
Jorgensen,~W.~L.; Maxwell,~D.; Tirado-Rives,~J. Development and Testing of the OPLS All-Atom Force Field on Conformational Energetics and Properties of Organic Liquids. \emph{Journal of the American Chemical Society} \textbf{1996}, \emph{118}, 11225--11236\relax
\mciteBstWouldAddEndPuncttrue
\mciteSetBstMidEndSepPunct{\mcitedefaultmidpunct}
{\mcitedefaultendpunct}{\mcitedefaultseppunct}\relax
\EndOfBibitem
\bibitem[Harder \latin{et~al.}(2015)Harder, Damm, Maple, Wu, Reboul, Xiang, Wang, Lupyan, Dahlgren, Knight, Kaus, Cerutti, Krilov, Jorgensen, Abel, and Friesner]{opls3}
Harder,~E. \latin{et~al.}  OPLS3: A Force Field Providing Broad Coverage of Drug-like Small Molecules and Proteins. \emph{Journal of chemical theory and computation} \textbf{2015}, \emph{12}, 281--296\relax
\mciteBstWouldAddEndPuncttrue
\mciteSetBstMidEndSepPunct{\mcitedefaultmidpunct}
{\mcitedefaultendpunct}{\mcitedefaultseppunct}\relax
\EndOfBibitem
\bibitem[Roos \latin{et~al.}(2019)Roos, Wu, Damm, Reboul, Stevenson, Lu, Dahlgren, Mondal, Chen, Wang, Abel, Friesner, and Harder]{opls3e}
Roos,~K.; Wu,~C.; Damm,~W.; Reboul,~M.; Stevenson,~J.~M.; Lu,~C.; Dahlgren,~M.~K.; Mondal,~S.; Chen,~W.; Wang,~L.; Abel,~R.; Friesner,~R.~A.; Harder,~E.~D. OPLS3e: Extending Force Field Coverage for Drug-Like Small Molecules. \emph{J Chem Theory Comput} \textbf{2019}, \emph{15}, 1863--1874\relax
\mciteBstWouldAddEndPuncttrue
\mciteSetBstMidEndSepPunct{\mcitedefaultmidpunct}
{\mcitedefaultendpunct}{\mcitedefaultseppunct}\relax
\EndOfBibitem
\bibitem[Lu \latin{et~al.}(2021)Lu, Wu, Ghoreishi, Chen, Wang, Damm, Ross, Dahlgren, Russell, Von~Bargen, Abel, Friesner, and Harder]{opls4}
Lu,~C.; Wu,~C.; Ghoreishi,~D.; Chen,~W.; Wang,~L.; Damm,~W.; Ross,~G.~A.; Dahlgren,~M.~K.; Russell,~E.; Von~Bargen,~C.~D.; Abel,~R.; Friesner,~R.~A.; Harder,~E. OPLS4: Improving Force Field Accuracy on Challenging Regimes of Chemical Space. \emph{Journal of chemical theory and computation} \textbf{2021}, \emph{17}, 4291--4300\relax
\mciteBstWouldAddEndPuncttrue
\mciteSetBstMidEndSepPunct{\mcitedefaultmidpunct}
{\mcitedefaultendpunct}{\mcitedefaultseppunct}\relax
\EndOfBibitem
\bibitem[Jakalian \latin{et~al.}(2000)Jakalian, Bush, Jack, and Bayly]{bcc1}
Jakalian,~A.; Bush,~B.~L.; Jack,~D.~B.; Bayly,~C.~I. Fast, efficient generation of high-quality atomic charges. AM1-BCC model: I. Method. \emph{Journal of Computational Chemistry} \textbf{2000}, \emph{21}, 132--146\relax
\mciteBstWouldAddEndPuncttrue
\mciteSetBstMidEndSepPunct{\mcitedefaultmidpunct}
{\mcitedefaultendpunct}{\mcitedefaultseppunct}\relax
\EndOfBibitem
\bibitem[Jakalian \latin{et~al.}(2002)Jakalian, Jack, and Bayly]{bcc2}
Jakalian,~A.; Jack,~D.~B.; Bayly,~C.~I. Fast, efficient generation of high-quality atomic charges. AM1-BCC model: II. Parameterization and validation. \emph{Journal of Computational Chemistry} \textbf{2002}, \emph{23}, 1623--1641\relax
\mciteBstWouldAddEndPuncttrue
\mciteSetBstMidEndSepPunct{\mcitedefaultmidpunct}
{\mcitedefaultendpunct}{\mcitedefaultseppunct}\relax
\EndOfBibitem
\bibitem[Bayly \latin{et~al.}(1993)Bayly, Cieplak, Cornell, and Kollman]{resp}
Bayly,~C.~I.; Cieplak,~P.; Cornell,~W.; Kollman,~P.~A. A well-behaved electrostatic potential based method using charge restraints for deriving atomic charges: the RESP model. \emph{The Journal of Physical Chemistry} \textbf{1993}, \emph{97}, 10269--10280\relax
\mciteBstWouldAddEndPuncttrue
\mciteSetBstMidEndSepPunct{\mcitedefaultmidpunct}
{\mcitedefaultendpunct}{\mcitedefaultseppunct}\relax
\EndOfBibitem
\bibitem[Zhang \latin{et~al.}(2023)Zhang, Liu, Zhang, Zhang, Cai, Bi, Du, Qin, Huang, Li, Shan, Zeng, Zhang, Liu, Li, Chang, Wang, Zhou, Liu, Luo, Wang, Jiang, Wu, Yang, Yang, Yang, Gong, Zhang, Shi, Dai, York, Liu, Zhu, Zhong, Lv, Cheng, Jia, Chen, Ke, E, Zhang, and Wang]{zhang2023dpa2}
Zhang,~D. \latin{et~al.}  DPA-2: Towards a universal large atomic model for molecular and material simulation. 2023\relax
\mciteBstWouldAddEndPuncttrue
\mciteSetBstMidEndSepPunct{\mcitedefaultmidpunct}
{\mcitedefaultendpunct}{\mcitedefaultseppunct}\relax
\EndOfBibitem
\bibitem[Wang \latin{et~al.}(2022)Wang, Fass, Kaminow, Herr, Rufa, Zhang, Pulido, Henry, Bruce~Macdonald, Takaba, and Chodera]{wang2022espaloma}
Wang,~Y.; Fass,~J.; Kaminow,~B.; Herr,~J.~E.; Rufa,~D.; Zhang,~I.; Pulido,~I.; Henry,~M.; Bruce~Macdonald,~H.~E.; Takaba,~K.; Chodera,~J.~D. End-to-end differentiable construction of molecular mechanics force fields. \emph{Chem Sci} \textbf{2022}, \emph{13}, 12016--12033\relax
\mciteBstWouldAddEndPuncttrue
\mciteSetBstMidEndSepPunct{\mcitedefaultmidpunct}
{\mcitedefaultendpunct}{\mcitedefaultseppunct}\relax
\EndOfBibitem
\bibitem[Lahey \latin{et~al.}(2020)Lahey, Thien~Phuc, and Rowley]{lahey2020anitorsion}
Lahey,~S.~J.; Thien~Phuc,~T.~N.; Rowley,~C.~N. Benchmarking Force Field and the ANI Neural Network Potentials for the Torsional Potential Energy Surface of Biaryl Drug Fragments. \emph{J Chem Inf Model} \textbf{2020}, \emph{60}, 6258--6268\relax
\mciteBstWouldAddEndPuncttrue
\mciteSetBstMidEndSepPunct{\mcitedefaultmidpunct}
{\mcitedefaultendpunct}{\mcitedefaultseppunct}\relax
\EndOfBibitem
\bibitem[Jacobson \latin{et~al.}(2022)Jacobson, Stevenson, Ramezanghorbani, Ghoreishi, Leswing, Harder, and Abel]{qrnn}
Jacobson,~L.~D.; Stevenson,~J.~M.; Ramezanghorbani,~F.; Ghoreishi,~D.; Leswing,~K.; Harder,~E.~D.; Abel,~R. Transferable Neural Network Potential Energy Surfaces for Closed-Shell Organic Molecules: Extension to Ions. \emph{J Chem Theory Comput} \textbf{2022}, \emph{18}, 2354--2366\relax
\mciteBstWouldAddEndPuncttrue
\mciteSetBstMidEndSepPunct{\mcitedefaultmidpunct}
{\mcitedefaultendpunct}{\mcitedefaultseppunct}\relax
\EndOfBibitem
\bibitem[Bannwarth \latin{et~al.}(2019)Bannwarth, Ehlert, and Grimme]{gfn2xtb}
Bannwarth,~C.; Ehlert,~S.; Grimme,~S. GFN2-xTB-An Accurate and Broadly Parametrized Self-Consistent Tight-Binding Quantum Chemical Method with Multipole Electrostatics and Density-Dependent Dispersion Contributions. \emph{J Chem Theory Comput} \textbf{2019}, \emph{15}, 1652--1671\relax
\mciteBstWouldAddEndPuncttrue
\mciteSetBstMidEndSepPunct{\mcitedefaultmidpunct}
{\mcitedefaultendpunct}{\mcitedefaultseppunct}\relax
\EndOfBibitem
\bibitem[smi()]{smirks}
SMIRKS. \url{https://www.daylight.com/dayhtml/doc/theory/theory.smirks.html}\relax
\mciteBstWouldAddEndPuncttrue
\mciteSetBstMidEndSepPunct{\mcitedefaultmidpunct}
{\mcitedefaultendpunct}{\mcitedefaultseppunct}\relax
\EndOfBibitem
\bibitem[Mobley \latin{et~al.}(2018)Mobley, Bannan, Rizzi, Bayly, Chodera, Lim, Lim, Beauchamp, Slochower, Shirts, Gilson, and Eastman]{smirnoff}
Mobley,~D.~L.; Bannan,~C.~C.; Rizzi,~A.; Bayly,~C.~I.; Chodera,~J.~D.; Lim,~V.~T.; Lim,~N.~M.; Beauchamp,~K.~A.; Slochower,~D.~R.; Shirts,~M.~R.; Gilson,~M.~K.; Eastman,~P.~K. Escaping Atom Types in Force Fields Using Direct Chemical Perception. \emph{Journal of Chemical Theory and Computation} \textbf{2018}, \emph{14}, 6076--6092, PMID: 30351006\relax
\mciteBstWouldAddEndPuncttrue
\mciteSetBstMidEndSepPunct{\mcitedefaultmidpunct}
{\mcitedefaultendpunct}{\mcitedefaultseppunct}\relax
\EndOfBibitem
\bibitem[Folmsbee and Hutchison(2020)Folmsbee, and Hutchison]{hutchison2020}
Folmsbee,~D.; Hutchison,~G. Assessing conformer energies using electronic structure and machine learning methods. \emph{International Journal of Quantum Chemistry} \textbf{2020}, \emph{121}\relax
\mciteBstWouldAddEndPuncttrue
\mciteSetBstMidEndSepPunct{\mcitedefaultmidpunct}
{\mcitedefaultendpunct}{\mcitedefaultseppunct}\relax
\EndOfBibitem
\bibitem[Stevenson \latin{et~al.}(2019)Stevenson, Jacobson, Zhao, Wu, Maple, Leswing, Harder, and Abel]{schrodinger-ani}
Stevenson,~J.; Jacobson,~L.~D.; Zhao,~Y.; Wu,~C.; Maple,~J.; Leswing,~K.; Harder,~E.; Abel,~R. Schrodinger-ANI: An Eight-Element Neural Network Interaction Potential with Greatly Expanded Coverage of Druglike Chemical Space. \textbf{2019}, \relax
\mciteBstWouldAddEndPunctfalse
\mciteSetBstMidEndSepPunct{\mcitedefaultmidpunct}
{}{\mcitedefaultseppunct}\relax
\EndOfBibitem
\bibitem[Wang \latin{et~al.}(2015)Wang, Wu, Deng, Kim, Pierce, Krilov, Lupyan, Robinson, Dahlgren, Greenwood, Romero, Masse, Knight, Steinbrecher, Beuming, Damm, Harder, Sherman, Brewer, Wester, A, Frye, Farid, Lin, Mobley, Jorgensen, Berne, Friesner, and Abel]{wang2015accurate}
Wang,~L. \latin{et~al.}  Accurate and Reliable Prediction of Relative Ligand Binding Potency in Prospective Drug Discovery by Way of a Modern Free-Energy Calculation Protocol and Force Field. \emph{Journal of the American Chemical Society} \textbf{2015}, \emph{137}, 2695--2703\relax
\mciteBstWouldAddEndPuncttrue
\mciteSetBstMidEndSepPunct{\mcitedefaultmidpunct}
{\mcitedefaultendpunct}{\mcitedefaultseppunct}\relax
\EndOfBibitem
\bibitem[Abraham \latin{et~al.}(2015)Abraham, Murtola, Schulz, Páll, Smith, Hess, and Lindahl]{gromacs}
Abraham,~M.; Murtola,~T.; Schulz,~R.; Páll,~S.; Smith,~J.~C.; Hess,~B.; Lindahl,~E. GROMACS: High performance molecular simulations through multi-level parallelism from laptops to supercomputers. \emph{SoftwareX} \textbf{2015}, \emph{1}, 19--25\relax
\mciteBstWouldAddEndPuncttrue
\mciteSetBstMidEndSepPunct{\mcitedefaultmidpunct}
{\mcitedefaultendpunct}{\mcitedefaultseppunct}\relax
\EndOfBibitem
\bibitem[Lindorff-Larsen \latin{et~al.}(2010)Lindorff-Larsen, Piana, Palmo, Maragakis, Klepeis, Dror, and Shaw]{99sb-ildn}
Lindorff-Larsen,~K.; Piana,~S.; Palmo,~K.; Maragakis,~P.; Klepeis,~J.~L.; Dror,~R.~O.; Shaw,~D.~E. Improved side-chain torsion potentials for the Amber ff99SB protein force field. \emph{Proteins: Structure, Function, and Bioinformatics} \textbf{2010}, \emph{78}, 1950--1958\relax
\mciteBstWouldAddEndPuncttrue
\mciteSetBstMidEndSepPunct{\mcitedefaultmidpunct}
{\mcitedefaultendpunct}{\mcitedefaultseppunct}\relax
\EndOfBibitem
\bibitem[Shirts and Chodera(2008)Shirts, and Chodera]{shirts2008mbar}
Shirts,~M.~R.; Chodera,~J.~D. Statistically optimal analysis of samples from multiple equilibrium states. \emph{The Journal of Chemical Physics} \textbf{2008}, \emph{129}\relax
\mciteBstWouldAddEndPuncttrue
\mciteSetBstMidEndSepPunct{\mcitedefaultmidpunct}
{\mcitedefaultendpunct}{\mcitedefaultseppunct}\relax
\EndOfBibitem
\bibitem[MacKerell~Jr \latin{et~al.}(2004)MacKerell~Jr, Feig, and Brooks]{cmap1}
MacKerell~Jr,~A.~D.; Feig,~M.; Brooks,~C.~L. Improved treatment of the protein backbone in empirical force fields. \emph{Journal of the American Chemical Society} \textbf{2004}, \emph{126}, 698--699\relax
\mciteBstWouldAddEndPuncttrue
\mciteSetBstMidEndSepPunct{\mcitedefaultmidpunct}
{\mcitedefaultendpunct}{\mcitedefaultseppunct}\relax
\EndOfBibitem
\bibitem[Mackerell~Jr. \latin{et~al.}(2004)Mackerell~Jr., Feig, and Brooks~III]{cmap2}
Mackerell~Jr.,~A.~D.; Feig,~M.; Brooks~III,~C.~L. Extending the treatment of backbone energetics in protein force fields: Limitations of gas-phase quantum mechanics in reproducing protein conformational distributions in molecular dynamics simulations. \emph{Journal of Computational Chemistry} \textbf{2004}, \emph{25}, 1400--1415\relax
\mciteBstWouldAddEndPuncttrue
\mciteSetBstMidEndSepPunct{\mcitedefaultmidpunct}
{\mcitedefaultendpunct}{\mcitedefaultseppunct}\relax
\EndOfBibitem
\bibitem[Eastman \latin{et~al.}(2023)Eastman, Behara, Dotson, Galvelis, Herr, Horton, Mao, Chodera, Pritchard, Wang, De~Fabritiis, and Markland]{spice}
Eastman,~P.; Behara,~P.~K.; Dotson,~D.~L.; Galvelis,~R.; Herr,~J.~E.; Horton,~J.~T.; Mao,~Y.; Chodera,~J.~D.; Pritchard,~B.~P.; Wang,~Y.; De~Fabritiis,~G.; Markland,~T.~E. SPICE, A Dataset of Drug-like Molecules and Peptides for Training Machine Learning Potentials. \emph{Sci Data} \textbf{2023}, \emph{10}, 11\relax
\mciteBstWouldAddEndPuncttrue
\mciteSetBstMidEndSepPunct{\mcitedefaultmidpunct}
{\mcitedefaultendpunct}{\mcitedefaultseppunct}\relax
\EndOfBibitem
\bibitem[Mardirossian and Head-Gordon(2016)Mardirossian, and Head-Gordon]{wb97m-v}
Mardirossian,~N.; Head-Gordon,~M. omegaB97M-V: A combinatorially optimized, range-separated hybrid, meta-GGA density functional with VV10 nonlocal correlation. \emph{J Chem Phys} \textbf{2016}, \emph{144}, 214110\relax
\mciteBstWouldAddEndPuncttrue
\mciteSetBstMidEndSepPunct{\mcitedefaultmidpunct}
{\mcitedefaultendpunct}{\mcitedefaultseppunct}\relax
\EndOfBibitem
\bibitem[Rappoport and Furche(2010)Rappoport, and Furche]{rappoport2010def2basis}
Rappoport,~D.; Furche,~F. Property-optimized gaussian basis sets for molecular response calculations. \emph{J Chem Phys} \textbf{2010}, \emph{133}, 134105\relax
\mciteBstWouldAddEndPuncttrue
\mciteSetBstMidEndSepPunct{\mcitedefaultmidpunct}
{\mcitedefaultendpunct}{\mcitedefaultseppunct}\relax
\EndOfBibitem
\bibitem[Gaulton \latin{et~al.}(2012)Gaulton, Bellis, Bento, Chambers, Davies, Hersey, Light, McGlinchey, Michalovich, Al-Lazikani, and Overington]{chembl2012}
Gaulton,~A.; Bellis,~L.~J.; Bento,~A.~P.; Chambers,~J.; Davies,~M.; Hersey,~A.; Light,~Y.; McGlinchey,~S.; Michalovich,~D.; Al-Lazikani,~B.; Overington,~J.~P. ChEMBL: a large-scale bioactivity database for drug discovery. \emph{Nucleic Acids Res} \textbf{2012}, \emph{40}, D1100--7\relax
\mciteBstWouldAddEndPuncttrue
\mciteSetBstMidEndSepPunct{\mcitedefaultmidpunct}
{\mcitedefaultendpunct}{\mcitedefaultseppunct}\relax
\EndOfBibitem
\bibitem[Zhang \latin{et~al.}(2020)Zhang, Wang, Chen, Zeng, Zhang, Wang, and E]{zhang2020dpgen}
Zhang,~Y.; Wang,~H.; Chen,~W.; Zeng,~J.; Zhang,~L.; Wang,~H.; E,~W. DP-GEN: A concurrent learning platform for the generation of reliable deep learning based potential energy models. \emph{Computer Physics Communications} \textbf{2020}, \emph{253}, 107206--NA\relax
\mciteBstWouldAddEndPuncttrue
\mciteSetBstMidEndSepPunct{\mcitedefaultmidpunct}
{\mcitedefaultendpunct}{\mcitedefaultseppunct}\relax
\EndOfBibitem
\bibitem[Chai and Head-Gordon(2008)Chai, and Head-Gordon]{wb97x-D}
Chai,~J.~D.; Head-Gordon,~M. Long-range corrected hybrid density functionals with damped atom-atom dispersion corrections. \emph{Phys Chem Chem Phys} \textbf{2008}, \emph{10}, 6615--20\relax
\mciteBstWouldAddEndPuncttrue
\mciteSetBstMidEndSepPunct{\mcitedefaultmidpunct}
{\mcitedefaultendpunct}{\mcitedefaultseppunct}\relax
\EndOfBibitem
\end{mcitethebibliography}
\providecommand{\latin}[1]{#1}
\makeatletter
\providecommand{\doi}
  {\begingroup\let\do\@makeother\dospecials
  \catcode`\{=1 \catcode`\}=2 \doi@aux}
\providecommand{\doi@aux}[1]{\endgroup\texttt{#1}}
\makeatother
\providecommand*\mcitethebibliography{\thebibliography}
\csname @ifundefined\endcsname{endmcitethebibliography}  {\let\endmcitethebibliography\endthebibliography}{}

\end{document}